\documentclass{ws-p8-50x6-00}

\begin{document}
\newcommand{\w}{wavelet}
\newcommand{\an}{analysis}

\title{STATUS OF RING-LIKE CORRELATIONS AND WAVELETS}

\author{I.M. Dremin}

\address{P.N.Lebedev Physical Institute, 117924 Moscow, Russia}

\maketitle

\abstracts{
The problem of large-scale correlations of particles produced in high-energy
collisions is discussed. Among them are, e.g., those correlations which lead to
ring-like and elliptic flow shapes of individual high-multiplicity events in
the polar+azimuthal angles plane. The \w\ method of \an\ which allows to
disentangle various patterns is proposed and applied to central Pb-Pb
collisions at 158AGeV.}

Various correlations of particles produced in high-energy collisions are known.
Especially well studied are the two-particle correlations due to the decay
of two-particle resonances and due to Bose-Einstein effect for identical
particles. The common correlation function technique is well suited for this
purpose. It is more difficult to apply it if several particles are involved.
Nevertheless, correlations leading to clusters (or mini-jets, non-reducible to
resonances) and to jets have been studied as well, however mostly in
$e^+e^-$-collisions where jet and subjet structures are clearly visible.
Otherwise one has to apply the averaging procedure to get knowledge of
dynamical correlations from studies of the moments of various distributions and
their behavior in different regions of the phase space (for recent reviews, see
\cite{dwdk,dgar}). It has lead to understanding of some global features 
of QCD \cite{dr93,ddre}.

However, the genuine multiparticle correlations originating from some
collective effects may appear in individual events.
There exists a method of the \w\ \an\ which allows to
recognize patterns due to correlations at different scales and damp down
the statistical noise even if it is very large in an initial sample.

Before delving into applications of this method, let us consider examples of
possible correlations which one would seek for. Everybody knows about Cherenkov
radiation of photons by a bunch of electrons traversing a medium whose
refractivity index exceeds 1. These photons form a ring in the plane
perpendicular to electrons motion, i.e. they are emitted at a definite polar
angle. As a hadronic analogue, one may treat an impinging nucleus as a bunch of
confined quarks each of which can emit gluons when traversing a target nucleus.
I speculated \cite{dr} about possible Cherenkov gluons relying
on experimental observation of the positive real part of the elastic forward
scattering amplitude of all hadronic processes at high energies. This is a
necessary condition for such process because in the commonly used formula for
the refractivity index its excess over 1 is proportional to this real part.
However, later \cite{dre2} I noticed that for such thin targets as nuclei the
similar effect can appear due to small confinement length thus giving us a new
tool for its estimate. If several gluons are emitted and each of them
generates a mini-jet, the ring-like substructure will be observed in the target
diagram. If the density of mini-jets within the ring is so high that they
overlap, then they form a ridge pattern (or a wall \cite{pesc}). If the number
of emitted gluons is not large, we'll see several jets (tower
structure \cite{pesc}) correlated in their polar, but not in the azimuthal
angle. Central collisions of nuclei are preferred for observation of
such effects because of a large number of participating partons. If the number
of correlated particles within the ring is large enough, it would result in
spikes in the
pseudorapidity distributions. However, the usual histogram method is not always
good to verify these spikes because it may split a single spike into two bins
thus diminishing its role. However, some hints to such structure can be found
from these histograms. The \w\ \an\ is well suited for this purpose because it
clearly resolves the local properties of a pattern on the event-by-event basis
by choosing the so-called Heisenberg windows (see, e.g., \cite{daub}).

Another example is provided by the elliptic flow i.e. the azimuthal
asymmetry in individual events. It may be related to a collective classical
sling-effect \cite{dman} of the rotation of colliding nuclei after peripheral
collisions initiated by the directed pressure \cite{olli} at some impact
parameter or to some jetty structures.It can tell us about the equation of state 
of the hadronic matter by the shape of the created squeezed states.
By passing, let me say that we also
observed elliptic flow patterns ("cucumber") corresponding to the large
value of the second Fourier coefficient, and the "three leaves flower"
pattern with large third coefficient but I do not discuss them here.

The event-by-event \an\ of patterns in experimental and Monte Carlo events
becomes especially important with the advent of RHIC and LHC. The 4$\pi $
acceptance of detectors will be crucial for the event-by-event \an\ .

When individual events are imaged visually, the human eye has a tendency
to observe different kinds of intricate patterns with dense clusters
(spikes) and rarefied voids. However, the observed effects are often dominated
by statistical fluctuations. The method of factorial moments was proposed
\cite{bpes} to remove the statistical background in a global \an\ and it shows
fractal properties even in the event-by-event approach.

Both in cosmic rays \cite{addk,alex,masl,arat,dtre} and at accelerators
\cite{maru,adam} some ring-like structures were observed in individual events.
Especially astonishing was the famous NA22 event of pion-nucleon interaction
with the pseudorapidity spike 60 times exceeding the average level.

Some event-by-event attempts (see, e.g., \cite{aaaa,cddh}) to treat nuclei
collisions by methods of the traditional inclusive and correlation measures
revealed that there are some jetty structures in individual events
which  lead to spikes in the angular (pseudorapidity) distribution. Various
Monte Carlo simulations of the process were compared to the data and failed
to describe this jettyness in its full strength. More detailed \an\
\cite{dlln,agab} of large statistics data on hadron-hadron interactions
(unfortunately, however, for rather low multiplicity) with
dense groups of particles well separated from other particles in an event
showed some "anomaly" in the angular distribution of these groups.
It was found that the centers of these groups prefer to be positioned at a definite
polar angle. This feature favors the above interpretation in terms of
Cherenkov gluons.

The \w\ \an\ reveals the
{\it local} properties of any pattern in an individual event at various scales
and, moreover, avoids smooth polynomial trends and underlines the fluctuation
patterns. By choosing the strongest fluctuations, one hopes to get those
dynamical ones which exceed the statistical component.

First attempts to use \w\ \an\ in multiparticle production go back to
P. Carruthers \cite{carr,lgca,gglc} who used \w s for diagonalisation of
covariance matrices of some simplified cascade models. The proposals of
correlation studies with the help of \w s were
promoted \cite{sboh,huan}, and used, in particular, for looking
for the disoriented chiral condensate \cite{shth,nand}. The
\w\ transform of the pseudorapidity spectra of JACEE events was done in 
\cite{sboh}.

The \w\ \an\ of patterns in central Pb-Pb collisions at 158AGeV
was first done in \cite{adk,dikk}. Five events with the highest registered
multiplicities from 1034 to 1221 charged particles were chosen from 
150 events of the emulsion
chamber experiment EMU15 at CERN by the group from Lebedev Physical Institute.

The target diagrams of secondary particles distributions for these events are
shown in Fig.1, where the radial distance from the center measures the polar
angle $\theta $, and the azimuthal angle $\phi $ is counted around the center.
One can sum over the azimuthal angle and plot the corresponding pseudorapidity
($\eta =-\log \tan \theta /2$) distributions shown in Fig.2. The pronounced
peaks ($\eta $-spikes) strongly exceeding expected
statistical fluctuations are seen in individual events. This inhomogeneity in
pseudorapidity can arise either due to a very strong jet i.e. a large group
(tower) of particles close both in polar and azimuthal angles or due to a
ring-like (ridge) structure when several jets with smaller number of particles
in each of them have similar polar angle but differ in their azimuthal angles.

In \cite{adk} \w s were first used to analyze two-dimensional
patterns of fluctuations of the event 19 in Fig.1. The results of the
one-dimensional analysis \cite{adk} of the two-dimensional target diagram are
presented. To proceed in this way, the whole azimuthal region was divided into
24 sectors. The pseudorapidity distributions in
each of them were separately analyzed. Neighboring sectors were
connected afterwards. Both jet and ring-like structures are found
from the values of squared wavelet coefficients as seen from Fig.3 
\cite{adk}. At small scales $a$, the wavelet analysis
reveals individual particles. At larger scales, the clusters or jets of
particles are resolved. Finally, at ever larger scale one notices the ring-like
structure around the center of the target diagram which penetrates from one
azimuthal sector to another at nearby values of the polar angle (pseudorapidity),
thus forming an elliptic ridge. This structure approximately corresponds to
the peak in the pseudorapidity distribution
(for more detail, see \cite{adk}). Let us note, that it is not easy to
notice in the target diagram of this event any increase of density
within the ring just by eye because of the specific properties of the
$\theta - \phi $ plot where the density of particles decreases fast toward
the external region of large polar angles.

To reveal these patterns in more detail one should perform the
two-dimensional local analysis. It is strongly desirable to get rid of such
drawback of the histogram method as fixed positions of bins that gives rise
to splitting of a jet into pieces contained in two or more bins. This chance
is provided by the wavelet transform of particle densities on the
two-dimensional plot. Wavelets choose automatically the size and shapes of
bins depending on particle densities at a given position.

In principle, the wavelet coefficients $W_{j_1,k_1,j_2,k_2}$ of the
two-dimensional function $f(\theta ,\phi )$ are found from the formula
\begin{equation}
W_{j_1,k_1,j_2,k_2}=\int f(\theta ,\phi )\psi (2^{-j_1}\theta -k_1;2^{-j_2}\phi
-k_2)d\theta d\phi .   \label{wav}
\end{equation}
Here $\theta _i, \phi _i$ are the polar and azimuthal angles of particles produced,
$f(\theta ,\phi )=\sum _i\delta (\theta -\theta _i)\delta (\phi -\phi _i)$
with a sum over all particles $i$ in a given event, $(k_1,k_2)$ denote the
locations and $(j_1,j_2)$ the scales analyzed. The function $\psi $ is the
analyzing wavelet. The higher the density fluctuations of particles in a given
region, the larger are the corresponding wavelet coefficients.

In practice, the discrete wavelets obtained from the tensor product of
two multiresolution analyses of standard one-dimensional Daubechies 8-tap
wavelets were used. Then the corresponding $ss, sd$ and $dd$ coefficients in
the two-dimensional matrix were calculated (see \cite{daub}). The common scale
$j_1=j_2=j$ was used. 

As stressed above, the ring-like structure should be a collective effect
involving many particles and large scales. Therefore,
to get rid of the low-scale background due to individual particles and analyze
their clusterization properties, we have chosen the scales $j>5$ where both
single jets and those clustered in ring-like structures can be revealed as
is seen from Fig.3. Therefore all coefficients
with $j<6$ are put equal to 0. The wavelet coefficients for any $j$
from the interval $6\leq j\leq 10$ are now presented as functions of
polar and azimuthal angles in a form of the two-dimensional landscape-like
surface over this plane i.e. over the target diagram. Their inverse wavelet
transform allows to get modified target diagrams of analyzed events with
large-scale structure left only. Higher fluctuations of particle density inside
large-scale formations and, consequently, larger wavelet coefficients
correspond to darker regions on this modified target
diagram shown in Fig.4. Here we demonstrate two events (numbered 3 and 6) from
those five shown in Figs.1 and 2. They display both jet and ring-like
structures which are different in different events. To discard
the methodical cut-off at $\eta \approx 1.6-1.8$ the region of $\eta >1.8$
was only considered.

Even though the statistics is very low, it was attempted to plot the
pseudorapidity distribution of the maxima of \w\ coefficients with the
hope to see if it reveals the peculiarities observed in high statistics but
low multiplicity hadron-hadron experiments \cite{dlln,agab}. In Fig. 5, the
number of highest maxima of wavelet coefficients exceeding the threshold value
$W_{j,k}>2\cdot 10^{-3}$ is plotted as a function of their pseudorapidity for
all five events considered. It is quite peculiar that positions of the maxima
are discrete. They are positioned quite symmetrically about the value
$\eta \approx 2.9$ corresponding to $90^0$ in cms as it should be for two Pb
nuclei colliding. Difference of heights is within the error bars.
More interesting, they do not fill in this central region but are rather
separated. Qualitatively, it coincides with findings in \cite{dlln,agab}.

For comparison, there were generated 100 central Pb-Pb interactions with energy
158AGeV according to Fritiof model and the same number of events
according to the random model describing the inclusive rapidity distribution
shape. The fluctuations in these events are much smaller
than in experimental ones and do not show any ring-like structure.

Thus I conclude that even on the qualitative level there is the noticeable
difference between experimental and simulated events with larger and somewhat
ordered fluctuations in the former ones. My aim here is to show the
applicability and power of the two-dimensional \w\ \an\ , the qualitative
features and
differences leaving aside quantitative characteristics till higher statistics
of high multiplicity AA-events becomes available. In particular, the special
automatic complex for emulsion processing with high space resolution in
Lebedev Physical Institute (www.lebedev.ru/structure/pavicom/index.htm)
is coming into operation, and it will enlarge the statistics of central Pb-Pb
collisions quite soon. STAR Collaboration of RHIC should also give soon some
data at ever higher energy. Wavelets provide a powerful tool for event-by-event
analysis of fluctuation patterns in such collisions.

\begin{center}
\vspace*{-3.2cm}
\epsfig{file=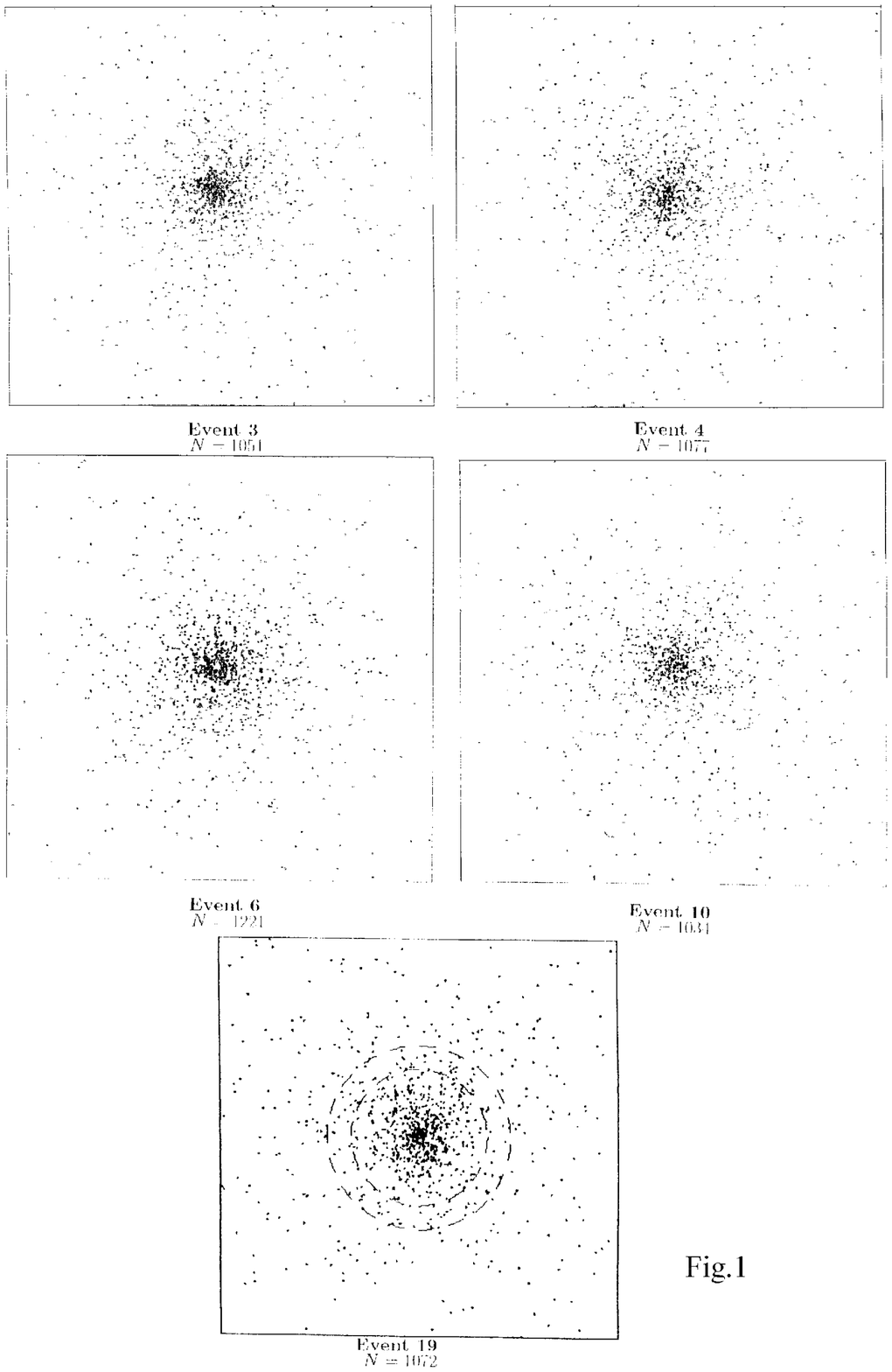,scale=0.65,clip=}
\end{center}

\vspace*{-2.8cm}
 The target diagrams of five events of central Pb-Pb collisions 
         at energy 158 GeV/nucleon obtained by EMU-15 collaboration.

\begin{center}
\vspace*{-2.5cm}
\epsfig{file=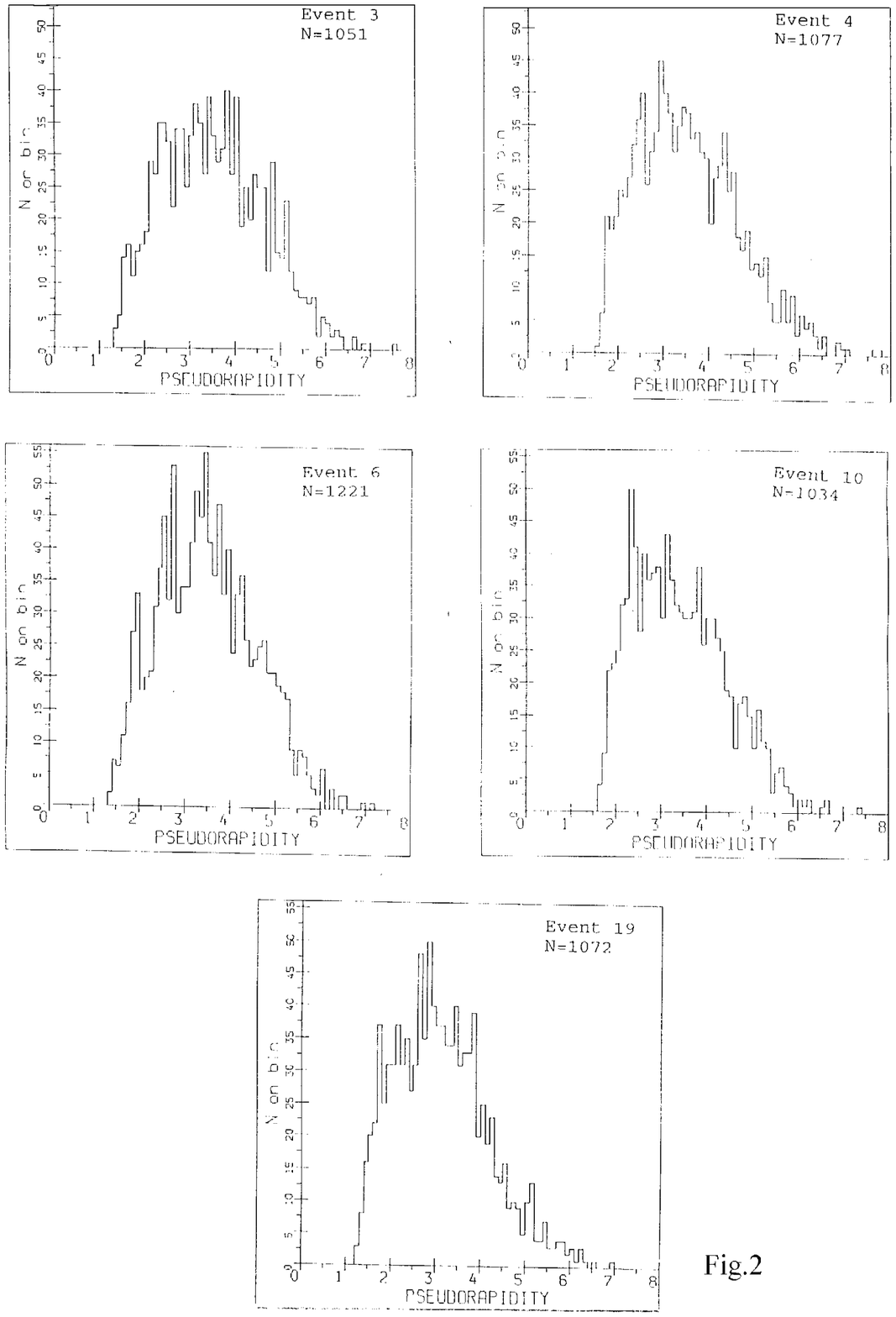,scale=0.7,clip=}
\end{center}
\vspace*{-1.5cm}
 The pseudorapidity distributions of particles in five events shown
         in Fig.1.\\

\begin{center}
\epsfig{file=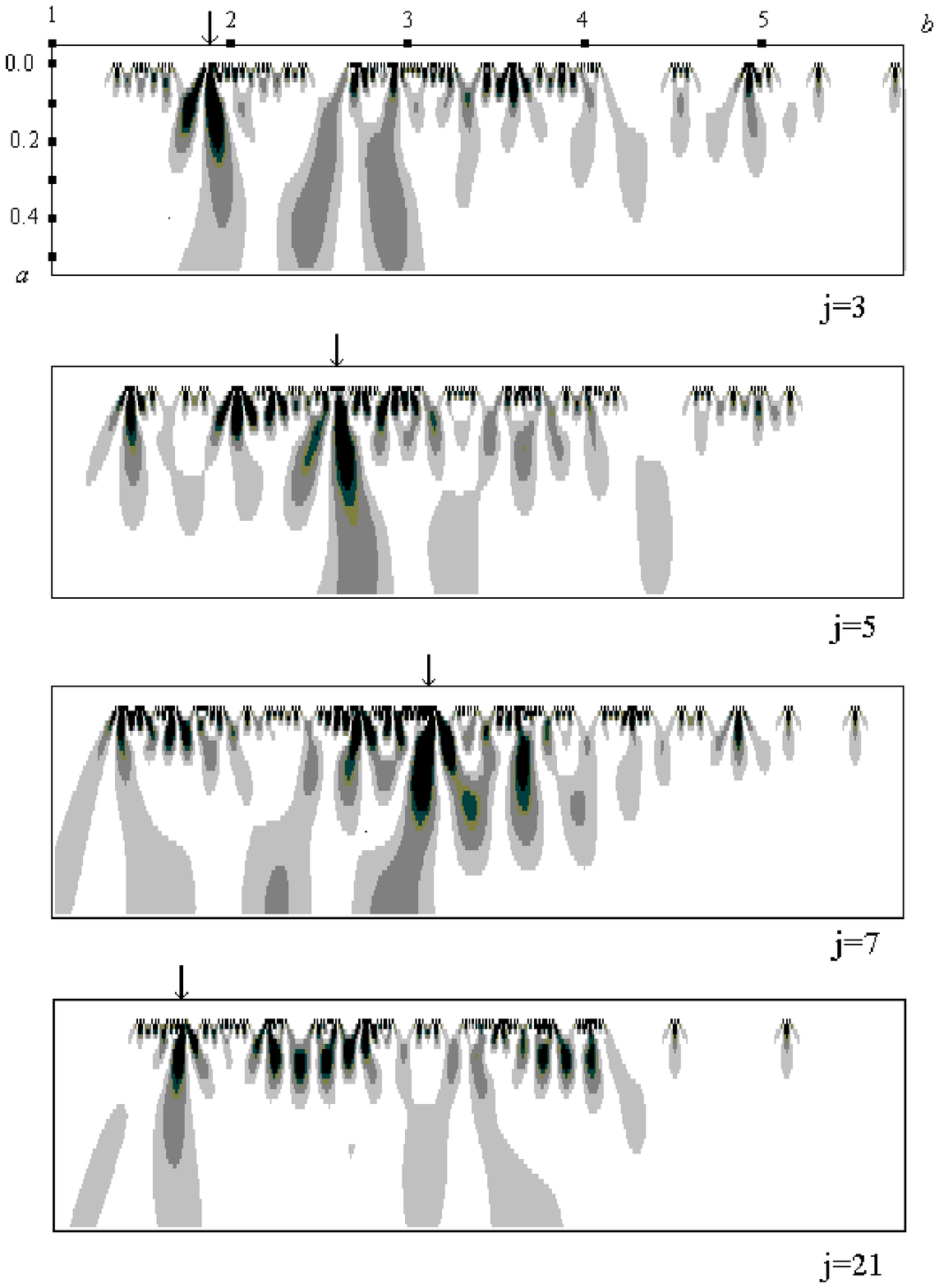, scale=0.6,clip=}
\end{center}
Fig. 3   The wavelet coefficients for the event 19 analyzed in \cite{adk}.
        Dark regions correspond to large values of the coefficients.      
         Four of 24 sectors are shown. Rapidities are along x-axis,
         the scales increase down the vertical axis.

\begin{center}
\begin{tabular}{cc}
\fbox{\epsfig{file=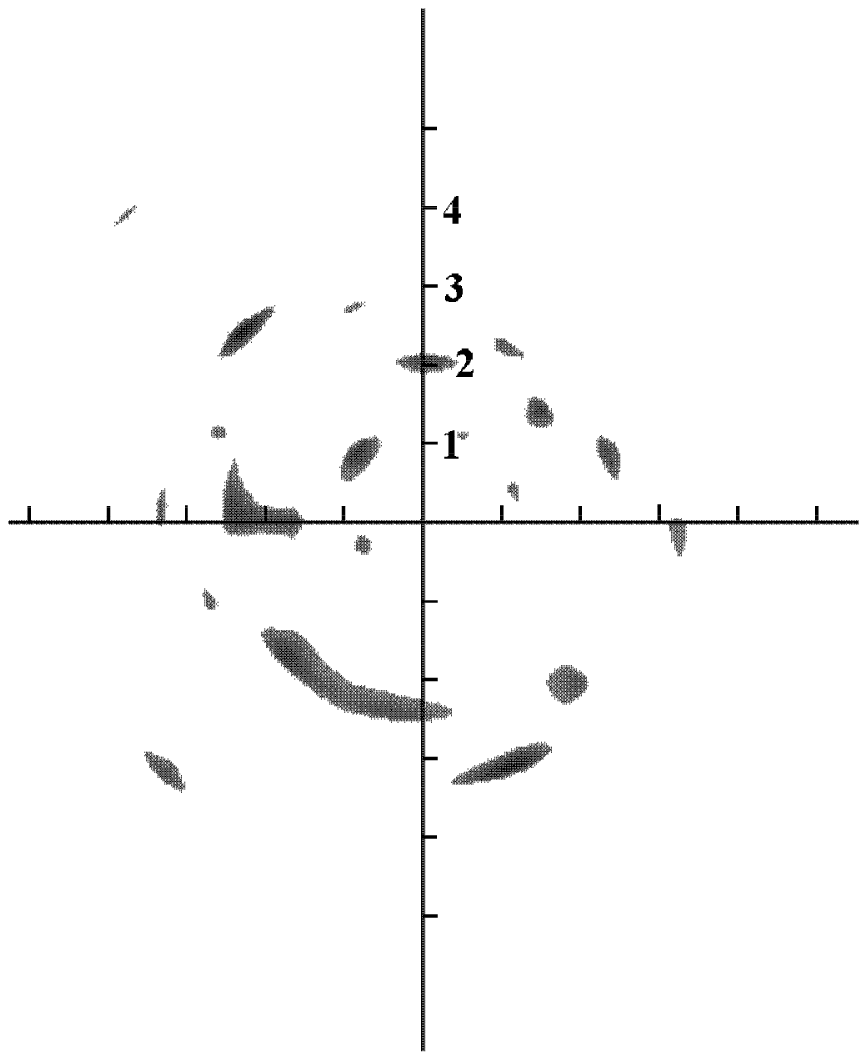,scale=0.47,clip=}}&
\fbox{\epsfig{file=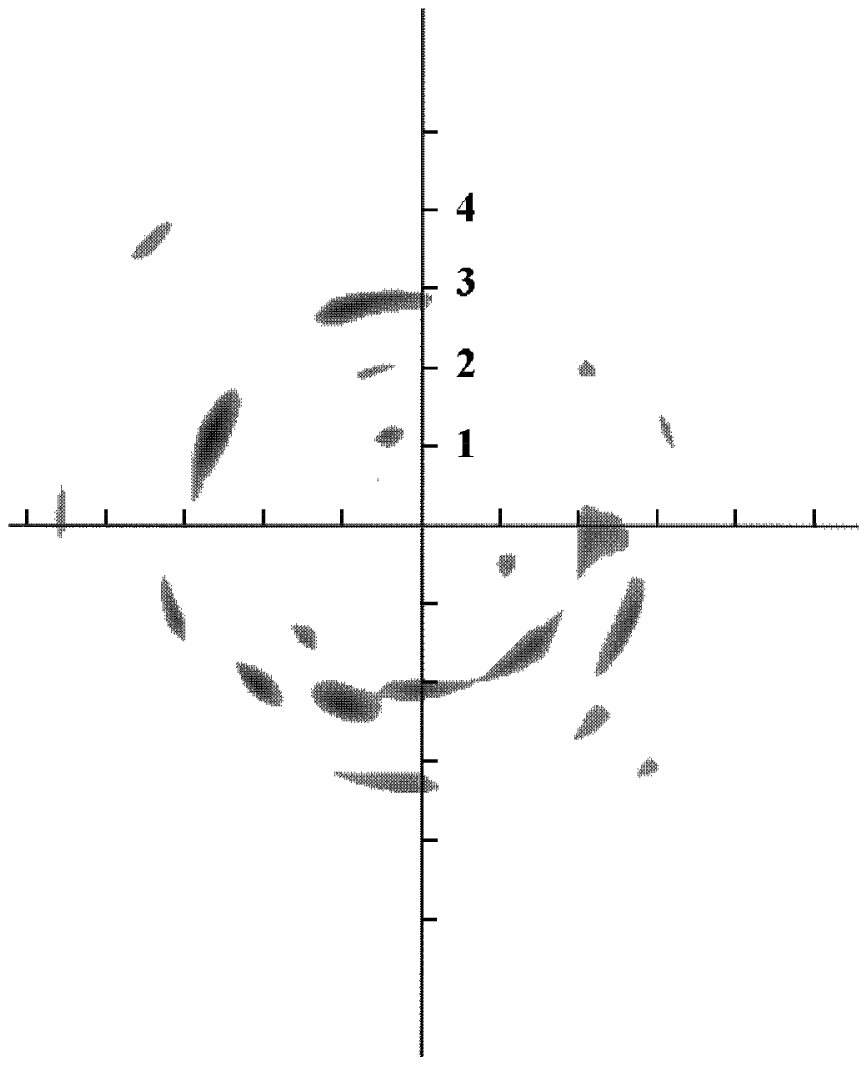,scale=0.47,clip=}}\
\end{tabular}
\end{center}
Fig.4 The modified large-scale target diagrams of two events (3 and 6).
      Darker regions correspond to larger particle density fluctuations. \\

\begin{center}
\fbox{\epsfig{file=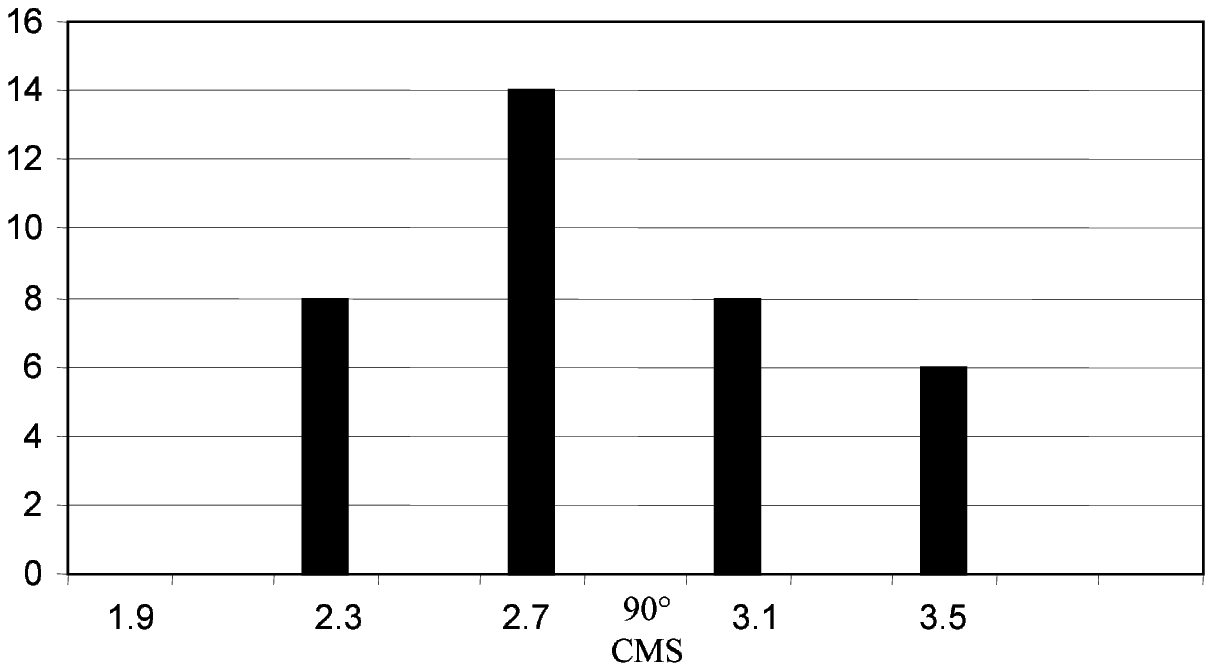,clip=}}
\end{center}
Fig.5 The pseudorapidity distribution of the maxima of \w\ coefficients.  \\
      The irregularity in the maxima positions, the empty voids between  \\
      them and absence of peaks at $\eta \approx 2.9$ are noticed. \\


\begin{thebibliography}{99}
\bibitem{dwdk}
E.A. De Wolf, I.M. Dremin and W. Kittel, Phys. Rep. 270 (1996) 1.
\bibitem{dgar}
I.M. Dremin and J.W. Gary, Phys. Rep. (2000) (to be published); hep-ph/0004215.
\bibitem{dr93}
I.M. Dremin, Phys. Lett. B313 (1993) 209.
\bibitem{ddre}
Yu.L. Dokshitzer and I.M. Dremin, Nucl. Phys. B402 (1993) 139.
\bibitem{dr}
I.M. Dremin, JETP Lett. 30 (1979) 140.
\bibitem{dre2}
I.M. Dremin, Yad. Fiz. 33 (1981) 1357.
\bibitem{pesc}
R. Peschanski, in Proc. XXII Int. Symp. on Multiparticle Dynamics (Santiago di
Compostella, 1992, Spain).
\bibitem{daub}
I. Daubechies, Ten Lectures on Wavelets. SIAM, Philadelphia, 1992.
\bibitem{dman}
I.M. Dremin and V.I. Manko, Nuovo Cim. A111 (1998) 439.
\bibitem{olli}
J.-Y. Ollitrault, Phys. Rev. D46 (1992) 229; D48 (1993) 1132.
\bibitem{bpes}
A. Bialas and R. Peschanski, Nucl. Phys. B273 (1988) 703.
\bibitem{addk}
A.V. Apanasenko, N.A. Dobrotin, I.M. Dremin et al, JETP Lett. 30 (1979) 145.
\bibitem{alex}
K.I. Alexeeva et al, Izvestia AN SSSR 26 (1962) 572; J. Phys. Soc. Japan 17,
A-III (1962) 409.
\bibitem{masl}
N.V. Maslennikova et al, Izvestia AN SSSR 36 (1972) 1696.
\bibitem{arat}
N. Arata, Nuovo Cim. 43A (1978) 455.
\bibitem{dtre}
I.M. Dremin, A.M. Orlov and M.I. Tretyakova, JETP Lett. 40 (1984) 320;
Proc. 17th ICRC v.5 (1981) 149.
\bibitem{maru}
I.A. Marutyan et al, Yad. Fiz. 29 (1979) 1566.
\bibitem{adam}
(NA22 Collaboration) M. Adamus et al, Phys. Lett. B185 (1987) 200.
\bibitem{aaaa}
EMU01 Collaboration, M.I. Adamovich et al, J. Phys. G19 (1993) 2035.
\bibitem{cddh}
KLM Collaboration, M.L. Cherry et al, Acta Physica Polonica B29 (1998) 2129.
\bibitem{dlln}
I.M. Dremin, P.L. Lasaeva, A.A. Loktionov et al, Sov. J. Nucl. Phys. 52
(1990) 840; Mod. Phys. Lett. A5 (1990) 1743.
\bibitem{agab}
(NA22 Collaboration) N.M. Agababyan et al, Phys. Lett. B389 (1996) 397.
\bibitem{carr}
P. Carruthers, in Proc. of "Hot and dense matter" Conference ( Bodrum, 1993), p.65.
\bibitem{lgca}
P. Lipa, M. Greiner and P. Carruthers, in Proc. of "Soft physics and fluctuations"
Conference (Krakow, 1993), p.105
\bibitem{gglc}
M. Greiner, J. Giesemann, P. Lipa and P. Carruthers, Z. Phys. C69 (1996) 305.
\bibitem{sboh}
N. Suzuki, M. Biyajima and A. Ohsawa, Prog. Theor. Phys. 94 (1995) 91.
\bibitem{huan}
D. Huang, Phys. Rev. D56 (1997) 3961.
\bibitem{shth}
I. Sarcevic, Z. Huang and R. Thews, Phys. Rev. D54 (1996) 750.
\bibitem{nand}
B.K. Nandi et al (WA98 Coll.), in Proc. of 3rd Int. Conference on Physics
and Astrophysics of Quark-Gluon Plasma (Jaipur, 1997), p.12.
\bibitem{adk}
N.M. Astafyeva, I.M. Dremin and K.A. Kotelnikov, Mod. Phys. Lett. A12 (1997)
1185.
\bibitem{dikk}
I.M. Dremin, O.V. Ivanov, S.A. Kalinin et al, Phys. Lett. B (2000) (to be
published).
\end{thebibliography}
\end{document}